\documentstyle[11pt,hakone]{article}
\input psfig
\def\msol{{\cal M}_\odot}
\def\ueber#1#2{{\setbox0=\hbox{$#1$}%
  \setbox1=\hbox to\wd0{\hss$ #2$\hss}%
  \offinterlineskip
  \vbox{\box1\box0}}{}}
\def\lesssim{\,\lower 1mm \hbox{\ueber{\sim}{<}}\,}
\def\grsim{\,\lower 1mm \hbox{\ueber{\sim}{>}}\,}
\begin{document}
\vspace*{13mm} \noindent \hspace*{13mm}
\begin{minipage}[t]{14cm}
{\bf X-RAY OBSERVATIONS OF DISTANT LENSING CLUSTERS}
\\[13mm]
S. Schindler\\
{\it Max-Planck-Institut f\"{u}r extraterrestrische Physik,
  Giessenbachstra\ss e, 85740, Garching, Germany}\\
{\it Max-Planck-Institut f\"{u}r Astrophysik,
 Karl-Schwarzschild-Stra\ss e 1, 85740, Garching, Germany}
\end{minipage}

\section*{Abstract}

X-ray observations of three clusters are presented: RXJ1347.5-1145,
Cl0939+47, and 
Cl0500-24. Although these clusters are the in same redshift range (0.32
- 0.45) and act all as gravitational lenses, 
they show very different properties.  RXJ1347.5-1145
seems to be an old, well relaxed system, with a relaxed morphology,
high X-ray luminosity, 
high temperature, high metallicity and strong cooling flow. The 
other two clusters have the
appearance of young systems with substructure and low X-ray
luminosity. The optical and X-ray luminosity
shows hardly any correlation. A comparison with nearby clusters shows
that many properties --
like e.g. the metallicity or the amount of subclustering
--  show a large scatter and no clear trend with time.

\section{Introduction}

Distant clusters are important objects to test cosmological models. A
comparison of the properties of distant clusters with the ones of
nearby clusters gives insight when and how the evolution took place.

Here we present the X-ray properties of three relatively distant
clusters in the redshift range of 0.32-0.45. 
All show a gravitational lensing effect. RXJ1347.5-1145 and
Cl0500-24 show bright arcs (Schindler et al. 1995; Giraud 1988),
Cl0939+47 shows a weak lensing signal (Seitz et al. 1996). 
The presence of a gravitational lensing signal means that they must be
all massive clusters. 

From these characteristics one might expect that they are similar in
other properties, too. But already the way how they are detected shows
that they are by far not similar. While Cl0939+47 and Cl0500-24 were
detected optically  (Cl0939+47 is actually the most distant Abell
cluster),  RXJ1347.5-1145 was detected in X-rays in the ROSAT All Sky
Survey. In the following we will show that also their X-ray properties
are very different.

\begin{figure}[!ht]
 \psfig{figure=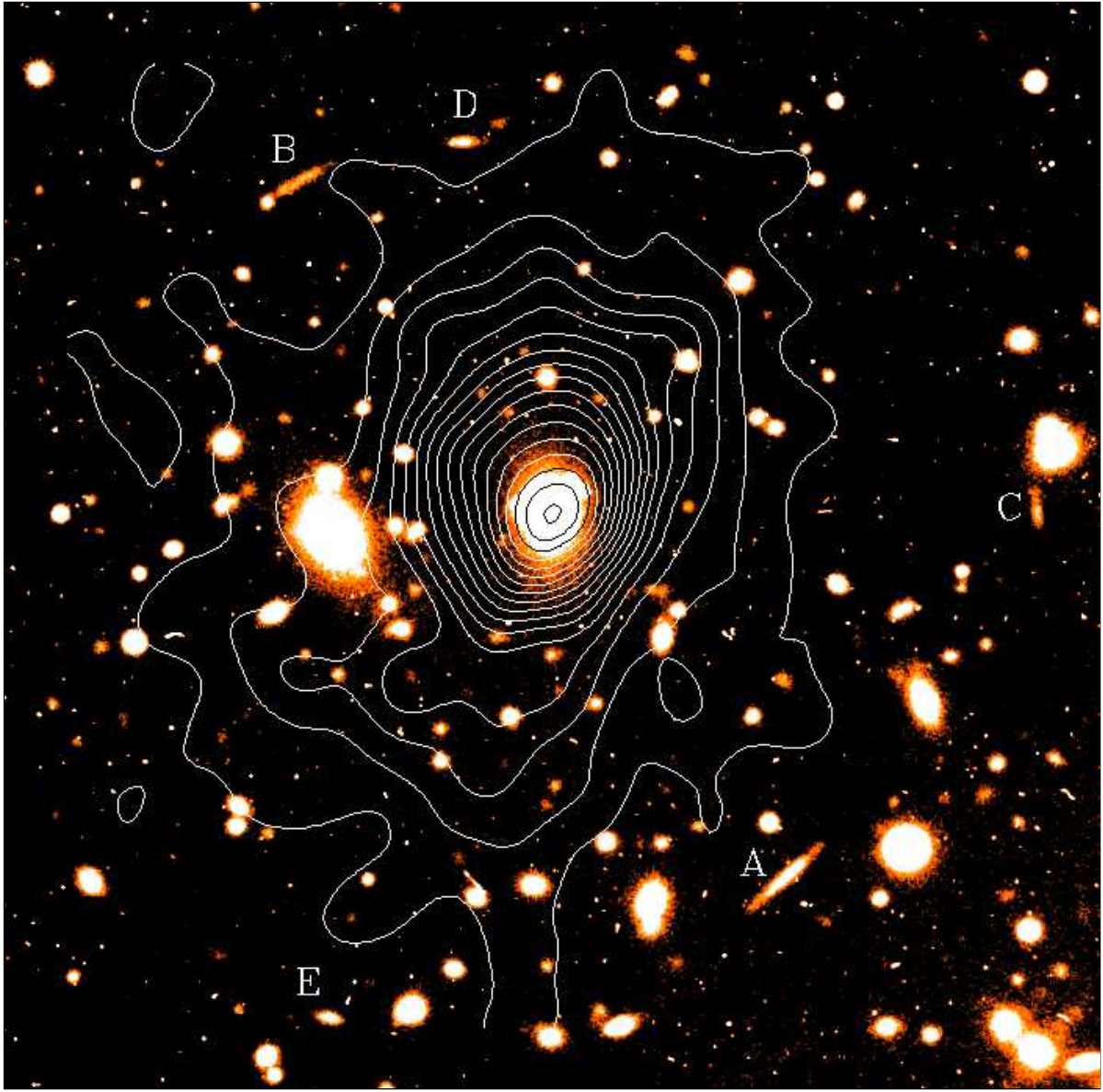,width=16.cm,clip=}
  \caption{ROSAT/HRI contours  of RXJ1347.5-1145
superposed on an R image taken at the NTT. The two images are
aligned in such a way that the positions of the X-ray maximum and the
central galaxy correspond.
The X-ray image is smoothed with a Gaussian filter of $\sigma$ = 2.5 arcsec.
The contours are linearly spaced with $\Delta$countrate =
0.032 counts/s/arcmin$^2$ the highest contour line corresponding to
0.54 counts/s/arcmin$^2$. The positions of the arcs and arc candidates
are marked with letters. 
The size of the image is $1.4\times1.4$arcmin$^2$ (North is
up, East is left).
}
\end{figure}

\section{The most luminous X-ray cluster RXJ1347.5-1145}

For the analysis of RXJ1347.5-1145 we use a ROSAT/HRI observation (see
Fig. 1) of 15760 seconds and an ASCA observation of 58300 seconds.
These data reveal several	 extreme cluster properties. 

The X-ray luminosity of RXJ1347.5-1145 is with 
$7.3\pm0.8\times 10^{45}$erg/s in the ROSAT
band (0.1-2.4 keV) or 
$2.1\pm0.4\times 10^{46}$erg/s bolometric the highest luminosity of a cluster
found so far. 

In the ASCA spectrum (Fig. 2) an Fe line can be detected. It
corresponds to a metallicity of $0.33\pm0.10$ in solar units. As this
is a typical value for nearby clusters, it is quite
surprising to find it in such a relatively distant cluster.
From the ASCA spectrum we can also determine the temperature. With
$9.3^{+1.1}_{-1.0}$ keV RXJ1347.5-1145 is a relatively hot cluster.

\begin{figure}[ht]
\psfig{figure=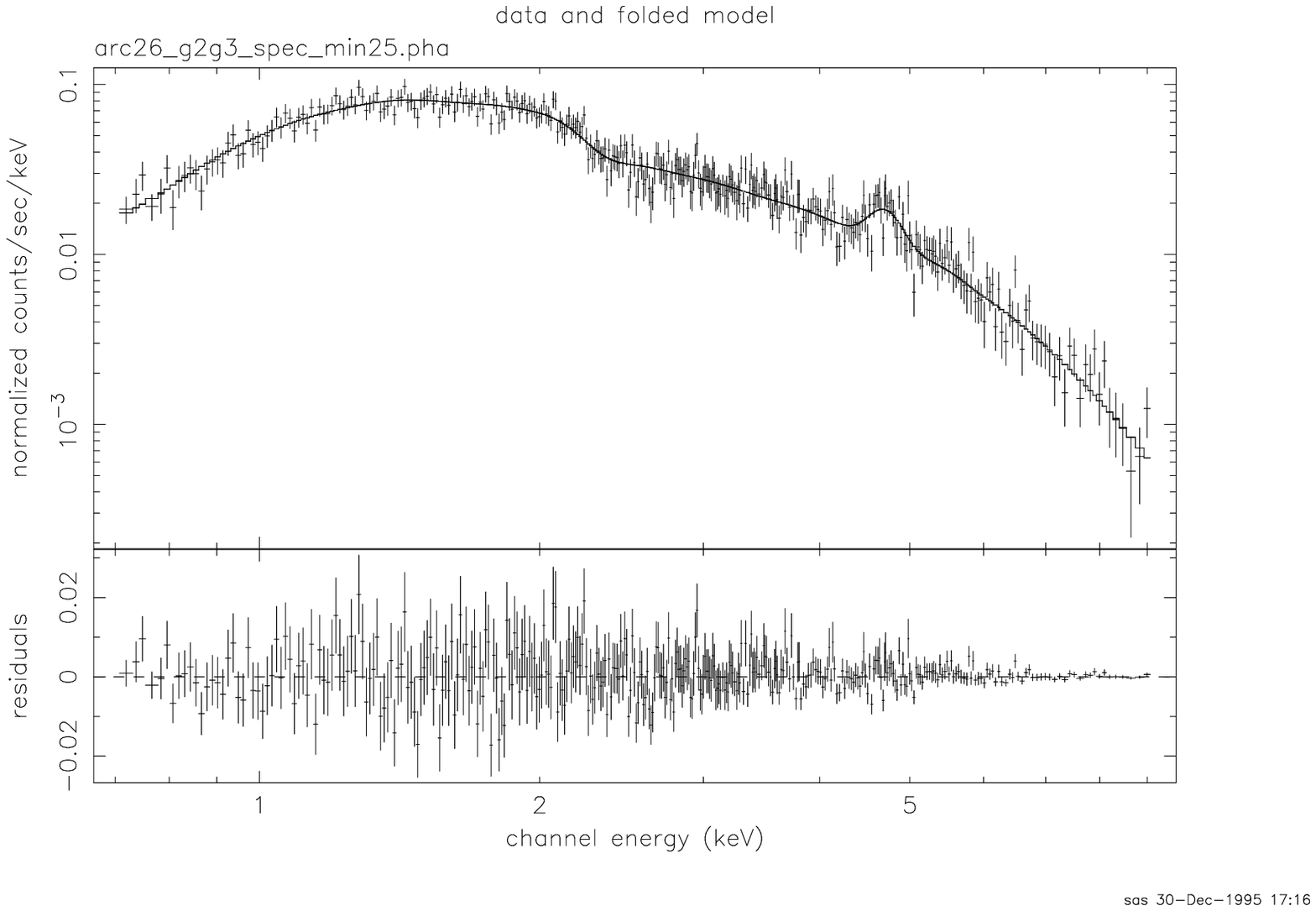,width=16.cm,clip=}
\caption{Spectrum of both GIS detectors within a 6.4 arcminute
radius. Around 5 keV the (redshifted) FeK line is visible, which
corresponds to a metallicity of 0.33 in solar units. A fit with a 
Raymond-Smith model (solid line) and the residuals are shown.}
\end{figure}

The strongly peaked emission (see Fig.1) suggests the presence of a
cooling flow. We find a central cooling time of $1.2\times10^9$
yr. With the standard assumptions we derive a cooling flow radius of
29 arcseconds (200 kpc) and a mass accretion rate of more than 3000
$\msol$/yr. Obviously, RXJ1347.5-1145 is also in terms of cooling
flow an extreme. Such a strong cooling flow suggests that was no
merging recently. Otherwise it would have disrupted the cooling flow
or at least decreased the mass accretion rate.

For a comparison of lensing and X-ray masses we calculate the surface
mass density from the X-ray data at the radius of the arcs,
$2.1\times 10^{14}\msol$. The lensing mass is still preliminary
because the redshift of the arcs are only estimated and the lens
model is very simple. For a redshift range of z=0.7-1.2, we find a
lensing mass of 4.4-7.8$\times10^{14}\msol$. This discrepancy can be
removed with a better lens model.

Summarizing, although RXJ1347.5-1145 is a 
distant cluster, it shows the properties of a well evolved, 
old system:  spherically symmetric morphology, high
luminosity, high temperature, high metallicity, and obviously no merging in the
recent past because of the huge cooling flow (see Table 1).
For more details see Schindler et al. (1997).

\section{The optically rich cluster Cl0939+47}

Cl0939+47 is an extremely rich, optically well studied cluster
(Dressler \& Gunn 1992). For the X-ray analysis we use a ROSAT/PSPC
observation of 14350 ksec.

Fig. 3 shows the ROSAT/PSPC image of
Cl0939+47. It has the appearance of a non-virialized cluster. 
It is not centrally peaked like RXJ1347.5-1145 but shows
substructure. There are three maxima visible. Ellipse fits to different
isophote levels yield ellipticities up to 0.75.

The X-ray luminosity is 
with $7.9\pm0.3 \times 10^{44}$ erg/s (0.1-2.4 keV) 
rather on the low side for such a rich cluster.
Also the temperature derived from the PSPC spectrum,
$2.9^{+1.3}_{-0.8}$ keV, is relatively low.

\begin{figure}[ht]
\psfig{figure=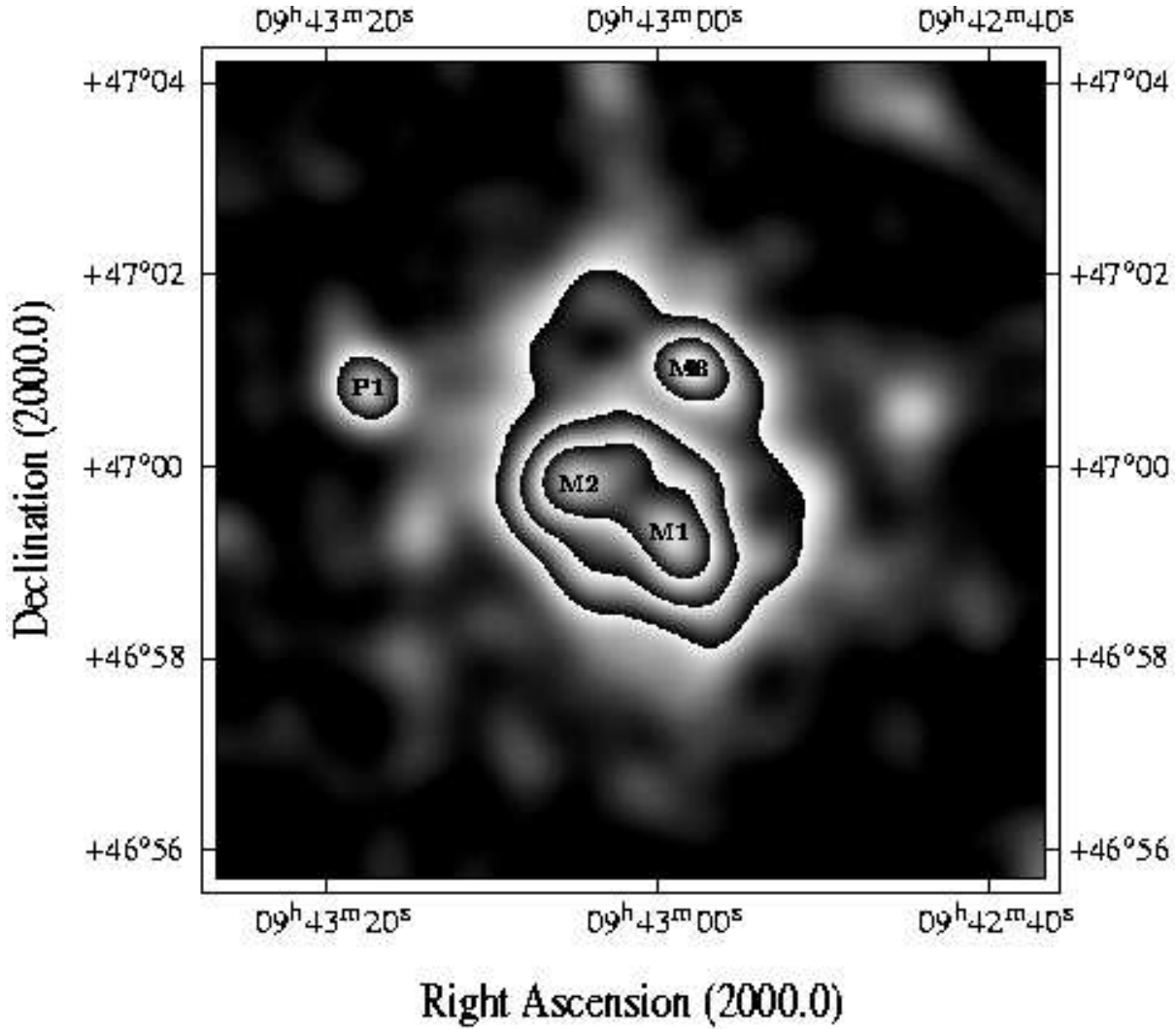,width=16.cm,clip=}
\caption{ROSAT/PSPC image of the cluster
CL0939+4713 in the ROSAT hard band (0.5-2.0 keV). 
The data are smoothed with a Gaussian filter of $\sigma$ = 15 arcsec.
Three maxima (M1, M2, M3) and a probable foreground point source (P1) 
are marked.  }
\end{figure}

A mass comparison is difficult for this cluster because, firstly, the 
weak lensing mass 
was determined only in the L-shaped region from an HST/WFPC
observation (Seitz et al. 1996) and the X-ray mass estimate has a
large error because spherical symmetry has to be assumed, which is not
a good approximation for this cluster. But it seems that the X-ray
mass is about a factor of three smaller than the lensing mass.

All the X-ray properties of Cl0939+47 (substructure, low X-ray
luminosity, low temperature) as well as the large fraction of
post-starburst galaxies (Belloni et al. 1995) point to a young,
non-relaxed system 
(for details see Schindler \& Wambsganss 1996).

\begin{figure}[!ht]
\psfig{figure=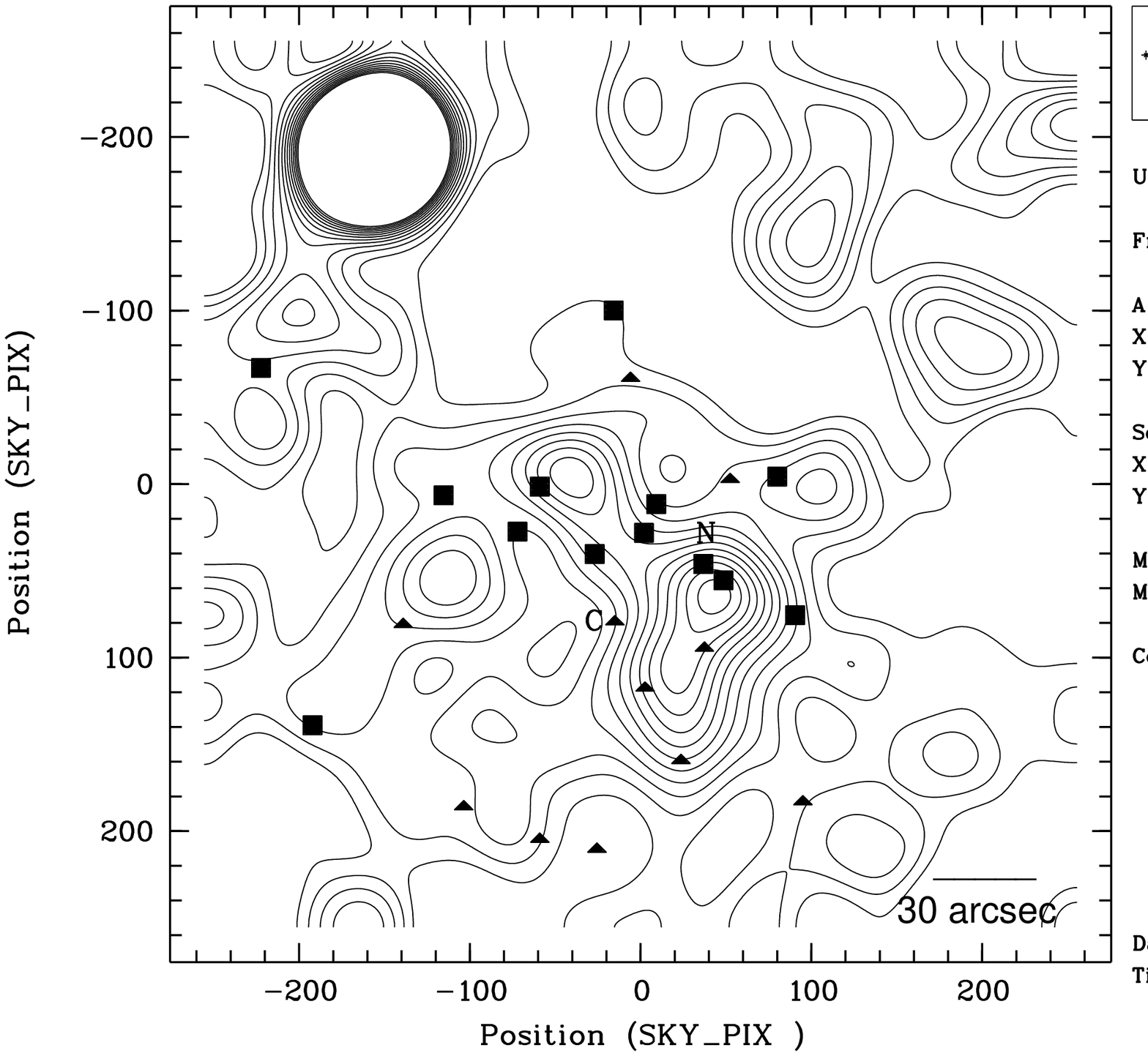,width=15.cm,clip=}
\caption{ROSAT/HRI contours of the cluster Cl0500-24 smoothed
with a Gaussian filter of $\sigma$ = 10 arcseconds.
The cluster has a clumpy structure with an extension south of the
maximum and additional emission in the north-east and in the east. The
bright X-ray source in the upper left corner is not associated with
the cluster. Superposed on the contours are
cluster galaxies assigned to the subclusters N and C by
Infante et al. (1994). Triangles:
galaxies assigned to subcluster C, squares: galaxies assigned to
subcluster N. One
tickmark corresponds to 10 arcseconds or 57 kpc. The
contour levels have a linear spacing of
$3.8\times10^{-4}$counts/s/arcmin$^2$. The highest contour corresponds
to $9.2\times10^{-3}$counts/s/arcmin$^2$, the lowest to 
$5.0\times10^{-3}$counts/s/arcmin$^2$. The central galaxies of the
subclusters are marked with N and C, respectively. The
X-ray emission is well correlated with the subcluster centred on N,
while there is hardly any correlation with the C subcluster. In
particular, there is no extra emission at the position of the central
galaxy C. }
\end{figure}

\section{Cl0500-24: a cluster with two subclusters in the line of sight}

The cluster Cl0500-24 is similar to Cl0939+47 in many respects. It is
also an optically rich cluster, which shows substructure. The
substructure is not only found in the ROSAT/HRI
image (Fig. 4), but also in the velocity distribution of the cluster
galaxies (Infante et al. 1994): they found two subclusters with a relative
velocity of about 3000 km/s.

A comparison of the spatial distribution of the subcluster galaxies
and the X-ray emission (Fig. 4) suggests that
only one of the subclusters is
X-ray luminous, the subcluster around galaxy N. This is an
indication that the N subcluster is massive. The C subcluster,
however, must be massive as well, because the arc has its curvature
towards galaxy C. Obviously, there are two components in this cluster
which have a very different gas content. 

The X-ray luminosity,
$3.1^{+0.6}_{-0.4}$ erg/s, is surprisingly low for such a rich
cluster. This fits well with the assumption that only part of the
cluster is X-ray luminous.

An X-ray  mass estimate yields a smaller mass than
the lensing mass model by Wambsganss et al. (1989). With the new ASCA
temperature (Ota et al. 1997; see also Mitsuda, this volume) 
the X-ray mass is $0.5\times10^{14}\msol$ at 22 arcmin
while the lensing model gives a mass of $1.4\times10^{14}\msol$ at the
same radius. This discrepancy
can be explained easily if one assumes that the cluster consists of
two subclusters, out of which only one is X-ray luminous. The X-ray
measurement traces only the potential well filled with gas, while lensing
is sensitive to all the mass along the line of sight. Furthermore, a
discrepancy can arise because the two mass estimates have different
centres: the X-ray mass is centred on the X-ray maximum (i.e. close to
galaxy N) while the mass
model is centred close to galaxy C.

Summarizing, Cl0500-24 shows the characteristics of a young system:
it has substructure and a low X-ray luminosity (for more details see 
Schindler \& Wambsganss 1997).

\begin{table}[!htb]
\begin{center}
\caption{Comparison of the X-ray properties of the three clusters. For
determining 
the X-ray mass of Cl0500-24 the ASCA temperature (Ota et
al. 1997; see also Mitsuda, this volume),
7.2 keV, is used.}
\begin{tabular}{|l|c|c|c|}
\hline
& & & \cr
       & RXJ1347.5-1145 & Cl0939+47 & Cl0500-24 \\
& & & \cr
\hline
redshift           & 0.45 & 0.41 & 0.32 \cr
$L_X$(0.1-2.4keV)[erg/s]  & $7.3\pm0.8\times10^{45}$
                   & $3.1\pm0.3\times10^{44}$
                   & $7.9^{+0.6}_{-0.4}\times10^{44}$ \cr
$r_c$ [kpc]        & 57 & 1100 & 30  \cr
$\beta$            & $0.56$&  1.9    & 0.36 \cr
metallicity (solar)& $0.33\pm0.10$  & - & -   \cr
$M_{gas}$($<1$Mpc)& 2.0$\times 10^{14}\msol$& 0.8$\times10^{14}\msol$& 
                     0.5$\times 10^{14}\msol$ \cr
$M_{tot}$($<1$Mpc)& $5.8\pm1.2\times 10^{14}\msol$&
                     $2.6^{+1.2}_{-0.6}\times10^{14}\msol$&
                     $2.7^{+1.4}_{-0.7}\times 10^{14}\msol$\cr
gas mass fraction ($<1$Mpc) & 30-40\% & 25-50\% & 12-25\%\cr
cooling flow radius& 200 kpc& - & -  \cr
central cooling time& $1.2\times10^9$yr & $3\times10^{10}$yr & - \cr
mass accretion rate& $\grsim 3000\msol$/yr  & - & -    \cr
\hline
\end{tabular}
\end{center}
\end{table}

\section{Conclusions}

Although the presented clusters are all massive and at about the same
distance, they are by far not similar. Some have the appearance of
young systems, still far away from virial equilibrium (Cl0939+47,
Cl0500-24), others seem to 
be already quite old systems (RXJ1347.5-1145). 
This difference is not only evident from 
the amount of substructure but also from the X-ray luminosity, the
gas temperature or the metallicity (see Tables 1 and 2).
These three clusters show hardly
any correlation of optical and X-ray luminosity.

\begin{table}[!htb]
\begin{center}
\caption{Comparison of nearby and distant clusters. $^a$ from Neumann
\& B\"ohringer (1997), $^b$ from Hattori et al. (1997), $^c$ from Ota et
al. (1997), see also Mitsuda
this volume, $^d$ from Tsuru et al. (1996), $^e$ from Neumann (1997)}
\begin{tabular}{|l|c|c|c|c|c|c|}
\hline
& & & & & & \cr
& nearby &Cl0500-24& Cl0939+47& RXJ1347&Cl0016+16$^a$&AXJ2019$^b$\cr
& & & & & & \cr
\hline
redshift &     & 0.32& 0.41 &0.45& 0.55 & 1.0       \cr

$L_X$(bol)[$10^{45}$erg/s]  & 0.05-5&0.6& 1.1 & 21 &5.0& 1.9    \cr
metallicity &0.2-0.5 & 0.0-1.5$^c$& small &  0.33  &small& $\approx 1.7$\cr
temperature & 2-10   & 7.2$^c$& 2.9 & 9.3 & 8.2$^d$ & 8.6  \cr 
substructure &in 25\%$^e$ & yes & yes & no  & yes & ?    \cr
\hline
\end{tabular}
\end{center}
\end{table}

A comparison of the properties of several distant clusters with
average properties of nearby clusters is shown in Table 2. Also the
clusters  Cl0016+16 (Neumann \& B\"ohringer 1997) and AXJ2019 (Hattori
et al. 1997) are included. The clusters are sorted according to their
distance. 
The comparison shows that there is a large scatter in the properties
but no clear evolutionary trend (see also Mushotzky \& Loewenstein
1997). This result points to a very early
evolution. Obviously, for studying evolutionary effects in clusters one has to
observe large samples of clusters, which are even more distant.





\bigskip It is a pleasure to thank Hans B\"ohringer, Makoto Hattori,
Doris Neumann, and
Joachim Wambsganss for inspiring collaborations. I
acknowledge financial support by the Verbundforschung.

\section*{References}
\begin{list}{}{\setlength{\leftmargin}{3em}\setlength{\rightmargin}{0cm}
        \setlength{\itemsep}{-0.4ex}\setlength{\baselineskip}{-0.7ex}
\setlength{\itemindent}{-3em}}

\item  Belloni, P., Bruzual, A.G., Thimm, G.J., R\"oser, H.-J.
	1995, A\&A, 297, 61
\item  Dressler, A., Gunn, J.E. 1992, ApJS,   78, 1
\item  Hattori, M., Ikebe, Y., Asaoka, I., Takeshima, T., B\"ohringer, H.,
       Mihara, T., Neumann, D.M., Schindler, S., Tsuru, T., Tamura,
       T. 1997, Nature, in press
\item  Giraud, E. 1988, ApJ,  334, L69 
\item  Infante, L., Fouqu\'e, P., Hertling, G., Way, M.J., Giraud, E.,
      Quintana H. 1994, A\&A, 289, 381
\item  Mushotzky, R.F. \& Loewenstein, M. 1997, ApJ, 481, L63 
\item  Neumann, D.M., B\"ohringer, H. 1997, MNRAS, 289, 123
\item   Neumann, D.M. 1997, Ph.D. Thesis,
Ludwigs-Maximilians-Universit\"at, M\"unchen
\item  Ota, N., Mitsuda, K., Fukazawa, Y., 1997, preprint
\item  Schindler, S.,  Guzzo, L., Ebeling, H., B\"ohringer, H., Chincarini, G.,
       Collins, C.A., De Grandi, S., Neumann, D.M., 
       Briel U.G., Shaver, P., Vettolani, G. 1995, A\&A, 299, L9
\item  Schindler, S., Wambsganss, J. 1996, A\&A, 313, 113
\item  Schindler, S., Wambsganss, J. 1997, A\&A, 322, 66
\item  Schindler, S., Hattori, M., Neumann, D.M., B\"ohringer, H. 1997,
          A\&A, 317, 646
\item  Seitz, C., Kneib J.-P., Schneider P., Seitz, S. 1996, A\&A,
          314, 707
\item  Tsuru, T., Koyama, K., Hughes, J.P., Arimoto, N., Kii, T., Hattori, M.
      1996, in UV and X-Ray Spectroscopy of Astrophysical and Laboratory
      Plasmas, ed.\ K. Yamashita K. and T. Watanabe (Tokyo: Universal Academic
      Press), 375
\item  Wambsganss, J., Giraud, E., Schneider, P., Weiss, A. 1989, ApJ, 337,
         L73

\end{list}
\end{document}